\documentclass[reprint,prb,amsmath,amssymb,amsfonts,showkeys]{revtex4-2}
\usepackage[dvipdfmx]{graphicx}
\usepackage{dcolumn}
\usepackage{bm}
\usepackage{times}
\usepackage{multirow}
\usepackage{color}

\usepackage{ulem}
\newcommand{\rc}[1]{\textcolor{black}{#1}}

\begin{document}

\title{\rc{Negatively enhanced thermopower near a Van Hove singularity in electron-doped Sr$_2$RuO$_4$}}

\author{Rei~Nishinakayama$^1$}
\email{6222526@ed.tus.ac.jp}
\author{Yoshiki~J.~Sato$^1$}
\thanks{\rc{Present address: Graduate School of Science and Engineering, Saitama University, Saitama 338-8570, Japan; yoshikisato@mail.saitama-u.ac.jp}}
\author{Takayoshi~Yamanaka$^2$}
\author{Yoshiteru~Maeno$^3$}
\author{Hiroshi~Yaguchi$^1$}
\author{Naoki~Kikugawa$^4$}
\author{Ryuji~Okazaki$^1$}
\email{okazaki@rs.tus.ac.jp}

\affiliation{
$^1$Department of Physics and Astronomy, Tokyo University of Science, Noda 278-8510, Japan\\
$^2$Institute for Materials Research, Tohoku University, Sendai 980-8577, Japan\\
$^3$Toyota Riken-Kyoto University Research Center (TRiKUC), Kyoto 606-8501, Japan\\
$^4$National Institute for Materials Science, Tsukuba 305-0003, Ibaraki, Japan}

\begin{abstract}

The layered perovskite Sr$_2$RuO$_4$ serves as a model material of the two-dimensional (2D) Fermi liquid
but also exhibits 
various emergent phenomena including the non-Fermi-liquid (NFL) behavior
under external perturbations
such as uniaxial pressure and chemical substitutions.
Here
we present the thermoelectric transport of electron-doped system Sr$_{2-y}$La$_{y}$RuO$_4$,
in which a filling-induced Lifshitz transition
occurs at the Van Hove singularity (VHS) point of $y\approx 0.2$.
We find that
the sign of the low-temperature thermopower becomes negative
only near the VHS point, where the NFL behavior has been observed in the earlier work.
This observation is
incompatible with either 
a numerical calculation
within a constant relaxation-time approximation
or
a toy-model calculation for the 2D Lifshitz transition
adopting an elastic carrier scattering.
\rc{As a promising origin of the observed negatively enhanced thermopower,}
we propose a skewed NFL state,
in which an inelastic scattering with a considerable odd-frequency term
plays a crucial role 
to negatively enhance the thermopower.

\end{abstract}

\maketitle

\section{Introduction}

Beyond the well-established Fermi-liquid (FL) picture,
non-Fermi-liquid (NFL) state has been intensively investigated
as an essential concept for
various 
quantum phenomena \cite{Hertz1976,Millis1993,Sachdev2008,Gegenwart2008,Stewart2001,Lohneysen2007,Chowdhury2022}.
Besides the well-known Tomonaga-Luttinger liquid in one dimension,
the prototypical NFL state appears in a vicinity of the 
quantum critical point (QCP),
in which 
a thermodynamic phase transition into an ordered state is 
suppressed by tuning external parameters. 
In the NFL state near QCP,
the finite-temperature properties such as 
the electronic specific heat and the electrical resistivity 
drastically deviate from the FL behavior,
as widely seen in 
correlated metals including
transition-metal oxides and heavy fermions.

A yet unsolved, fundamental issue
is 
how the FL picture is modified
in a vicinity of the Lifshitz transition \cite{Lifshitz1960},
an electronic topological transition 
associated with the change in the topology of the Fermi surfaces.
The Lifshitz transition itself is ubiquitous;
it is driven by 
various parameters such as pressure \cite{Nishimura2019,Krottenmuller2020,Sen2020}, 
magnetic field \cite{Pfau2017,Grachtrup2017,Pourret2019,Wu2023}, 
band filling \cite{Okamoto2010,Kwon2019,Ito2019}, 
and even temperature 
through the temperature-dependent chemical potential \cite{Wu2015,Zhang2017,Chen2020,Tian2021}.
In spite of many observed examples,
the kinetic properties near the Lifshitz transition
are complex and still controversial
owing to 
a peculiar energy-dependent relaxation time \cite{Varlamov1989,Varlamov2021},
which should become more complicated in correlated metals.

To tackle this problem, 
here we focus on 
the quasi-two-dimensional (q-2D) FL material 
Sr$_2$RuO$_4$ \cite{Maeno1994,Mackenzie2003,Mackenzie2017,Kivelson2020}.
Indeed,
its q-2D Fermi surfaces
consisting of hole-like $\alpha$ and electron-like $\beta$ and $\gamma$ sheets 
have been accurately verified by the dHvA and the ARPES experiments 
\cite{Mackenzie1996,Damascelli2000,Shen2007,Tamai2019}.
The calculated Fermi surfaces of Sr$_2$RuO$_4$ are shown in Fig. 1(a).
The normal-state nature in such a q-2D multi-band system
is well understood within the FL picture \cite{Mackenzie2003,Stricker2014,Bergemann2003},
whereas the pairing mechanism of the 
superconducting state is still an unsolved issue \cite{Pustogow2019,Ishida2020,Benhabib2021,Ghosh2021,Grinenko2021}.
Significantly, 
its superconducting transition temperature $T_{\rm c}$ 
is enhanced to $T_{\rm c} \approx 3.5$~K
under compressional stress \cite{Hicks2014,Taniguchi2015,Steppke2017,Li2021},
and 
near the
critical compression point in which $T_{\rm c}$ has a maximum value.
Such variation of $T_{\rm c}$ corresponds to the sharp peak in the density of states (DOS)
associated with a Van Hove singularity (VHS) point;
such a topology change in the $\gamma$ band is indeed observed by the 
ARPES \cite{Sunko2019}.
Most importantly, 
the resistivity clearly  
deviates from the FL behavior near the VHS point \cite{Barber2018,Herman2019}, 
and also
nontrivial electronic states such as
entropic anomaly \cite{Li2022} have been observed near the Lifshitz transition \cite{Luo2019,Morales2023,Kim2023}.
Thus, this layered material offers a suitable platform to investigate 
the NFL nature near the Lifshitz transition.

In this paper, we report 
a thermopower study of 
the electron-doped system Sr$_{2-y}$La$_y$RuO$_4$ \cite{Kikugawa2004_La,Kikugawa2004_Hall,Kikugawa2004},
in which a filling-induced Lifshitz transition occurs near the critical concentration $y_{\rm c}\approx 0.2$ \cite{Shen2007}.
The calculated Fermi surfaces for $y=y_{\rm c}$ and $y>y_{\rm c}$ are depicted in Figs. 1(b) and 1(c), respectively;
the topology of the $\gamma$ sheet changes at $y=y_c$.
In Sr$_{2-y}$La$_{y}$RuO$_4$, similar electronic features including NFL transport \cite{Kikugawa2004_La} and the 
effective mass enhancement \cite{Shen2007} have been clearly observed near $y = 0.2$,
offering a complementary approach toward such an intriguing issue
on the Lifshitz transition.
We find that
the low-temperature thermopower depends on the La content $y$ and that
the sign of the thermopower becomes negative only near the VHS point.
This is in contrast to the results of the Hall effect measurements
in which the Hall coefficient exhibits no significant anomaly near $y_{\rm c}$ \cite{Kikugawa2004_Hall}.
We also show that
the present experimental results
cannot be explained either by
numerical calculation results
within a constant relaxation-time approximation
or by a simple model for the 2D neck-disruption-type Lifshitz transition
with an elastic carrier scattering.
Instead,
we propose a skewed NFL state \cite{Georges2021} \rc{as a promising explanation for our results},
in which an odd-frequency inelastic scattering is considered.
The skewed NFL state is indeed a unique state of matter, as it has been studied as a nature of strange metal in cuprate superconductors and may also have a relevance to the transport properties in twisted bilayer graphene \cite{Georges2021}.
This skewed NFL state strengthens the electron-hole asymmetry owing to the dominant odd-frequency term in the scattering rate,
providing a crucial role for negatively enhanced thermopower near the Lifshitz transition.

\section{Experimental}

Single crystals of Sr$_{2-y}$La$_{y}$RuO$_4$ were grown by a floating-zone method \cite{Kikugawa2004_La,Kikugawa2004_Hall,Kikugawa2004}.
Typical dimension of the single crystals is $3\times 1\times 0.1$~mm$^3$.
The in-plane thermopower was measured by a steady-state technique using a manganin-constantan 
differential thermocouple in a closed-cycle refrigerator \cite{Kouda2022,Yamanaka2022}. 
A typical temperature gradient of 0.5 K/mm, which is adjusted along with the bath temperature, was applied along 
the in-plane direction using a resistive heater
and the distance between the thermocouple contacts is about 1 mm.
The thermoelectric voltage from the wire leads was subtracted.

\begin{figure}[t]
\begin{center}
\includegraphics[width=8cm]{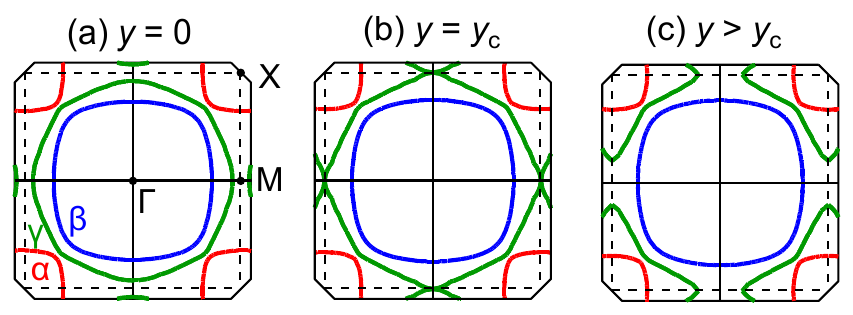}
\caption{
The cross-sectional view of the calculated Fermi surfaces of Sr$_{2-y}$La$_{y}$RuO$_4$ 
at $k_z=0$ plane for (a) $y=0$, (b) $y=y_{\rm c}$, and (c) $y>y_{\rm c}$, 
drawn by using FermiSurfer program \cite{Kawamura2019}.
}
\end{center}
\end{figure}

\section{Results and discussion}

Figure 2(a) shows the 
temperature dependence of the in-plane thermopower $S$ of Sr$_{2-y}$La$_{y}$RuO$_4$.
It is known that $T_{\rm c}$ decreases with La substitution and is completely suppressed for $y$ greater than 0.04 \cite{Kikugawa2004_La}.
For the parent compound, 
the present results well agree with those of the thermopower in previous reports \cite{Yoshino1996,Xu2008,Yamanaka2022,Daou2023,Otsuki2023}.
The thermopower of Sr$_2$RuO$_4$ has also been studied by the dynamical mean-field theory \cite{Mravlje2016}.
In the La-substituted compounds, 
overall behavior of the thermopower 
is similar to that of the parent crystal.

\begin{figure}[t!]
\begin{center}
\includegraphics[width=8cm]{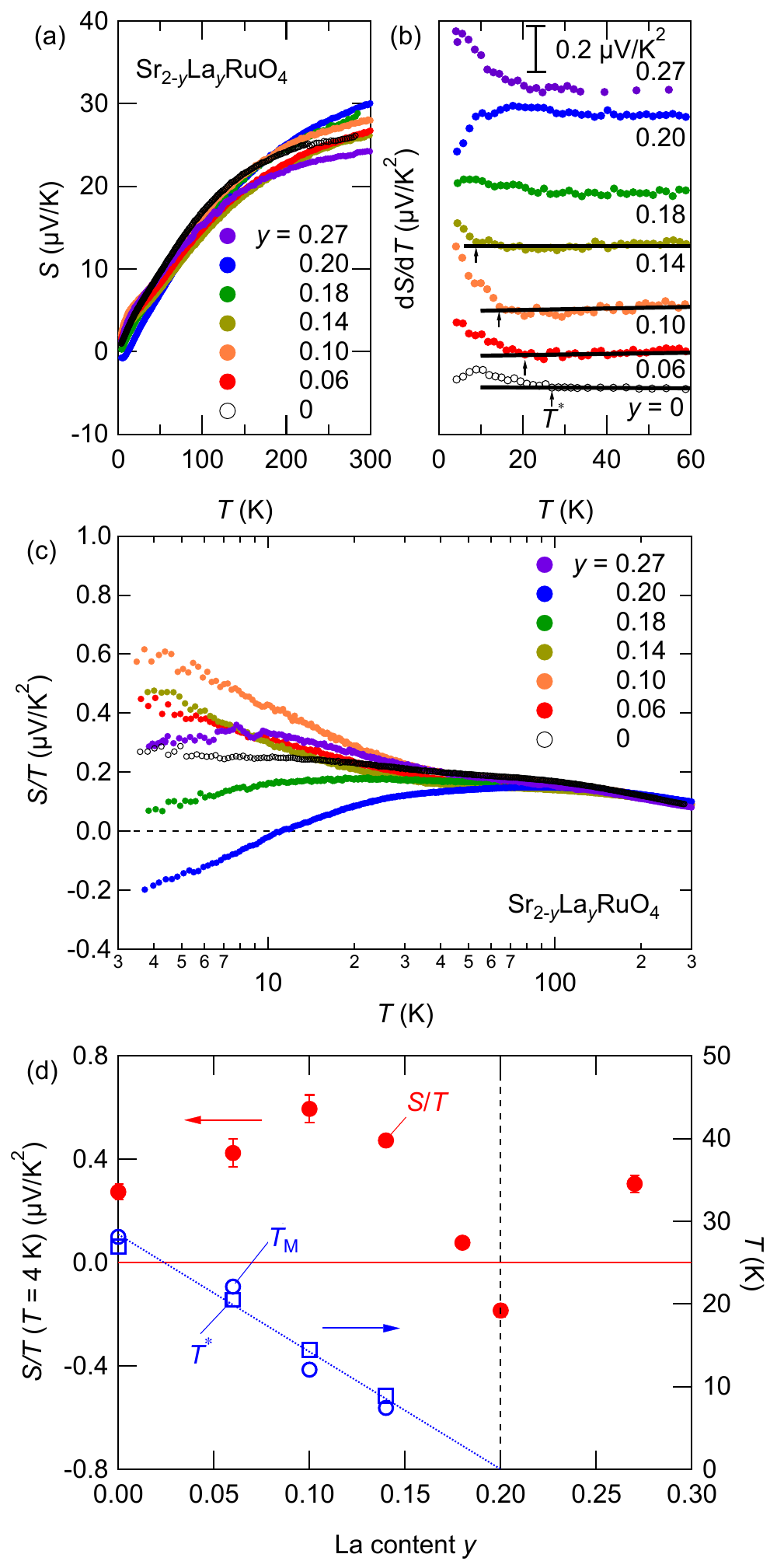}
\caption{
(a)
Temperature dependence of the thermopower $S$ of Sr$_{2-y}$La$_{y}$RuO$_4$ $(0\leq y \leq 0.27)$ measured along the $ab$ plane direction. 
(b) Temperature dependence of $dS/dT$.
The arrows show the characteristic temperature $T^*$
below which the $dS/dT$ deviates from the linear temperature dependence of 
$dS/dT$ at higher temperatures.
The data are shifted vertically for clarity.
(c)
Temperature dependence of $S/T$.
(d)
The La content $y$ dependence of $S/T$ measured at 4~K (solid circles, left axis),
$T^*$ (open squares, right axis), and $T_{\rm M}$ (open circles, right axis).
$T_{\rm M}$ is defined at the peak temperature in the 
magnetic susceptibility \cite{Kikugawa2004}. 
The blue dotted line is a guide to the eye to represent the $y$ dependence of these characteristic temperatures.
The vertical dashed line show the critical La content $y_{\rm c}$. 
}
\end{center}
\end{figure}

The thermopower in Sr$_2$RuO$_4$ was analyzed 
in the differential form $dS/dT$ [Fig. 2(b)] to examine a
characteristic temperature,
and an anomaly was found near $T^*\approx 25$~K,
below which  $dS/dT$ increases with decreasing temperature \cite{Yoshino1996}.
Subsequently, 
through
the Seebeck and the Nernst measurements,
Xu \textit{et al.} have
suggested that 
the coherence is developed below $T^*$ \cite{Xu2008}.
As displayed in the right axis of Fig. 2(d),
$T^*$ systematically decreases with the La content $y$.
It should be noted that
the magnetic susceptibility of Sr$_2$RuO$_4$ is Pauli paramagnetic, but
the temperature dependence exhibits a small peak structure at $T_{\rm M}\approx 30$~K \cite{Mackenzie2003},
below which the FL picture is well defined.
The La content $y$ dependence of $T_{\rm M}$ taken from Ref. \cite{Kikugawa2004} is also plotted in the right axis of Fig. 2(d).
Notably, 
both $T_{\rm M}$ and $T^*$ show similar $y$ dependence,
indicating that 
the characteristic temperature below which 
the coherence is formed in the correlated carriers 
decreases with increasing $y$.
This trend is consistent with the NFL behavior near $y=0.2$ \cite{Kikugawa2004_La}
and also signifies the inherent role of the carrier scattering at the Lifshitz transition, as will be discussed later.

In Fig. 2(c), 
we also plot $S/T$ of Sr$_{2-y}$La$_{y}$RuO$_4$ as a function of $T$.
The low-temperature $S/T$ behavior notably depends on both temperature and the La content $y$,
suggesting the considerable change in the Fermi surfaces in the present La content range 
as reported in earlier reports \cite{Kikugawa2004_La,Kikugawa2004_Hall,Kikugawa2004,Shen2007}.
It should be noted that 
negative thermopower is found at low temperatures only for the $y=0.2$ crystal,
which is near the critical La content $y_{\rm c}$ \cite{Kikugawa2004_La,Kikugawa2004_Hall,Kikugawa2004,Shen2007}.
Figure 2(d) displays the La content $y$ dependence of $S/T$ obtained at 4~K for the left axis.
With increasing $y$, $S/T$ slightly increases and 
a singularity is clearly observed near $y \approx 0.2$.
Note that a positive value of $S/T$ is recovered at the higher content $y=0.27$.
To explain the observed La content dependence of $S/T$,
we have examined the chemical potential dependence of
$S/T$ calculated within a constant relaxation-time approximation \cite{Madsen2006} (Appendix),
but the calculated data [Fig. S2(b)] is positively enhanced near the VHS points.
Obviously, this discrepancy originates from the energy dependence of the relaxation time,
which is ignored in the constant relaxation-time approximation method.

\begin{figure}[t]
\begin{center}
\includegraphics[width=8.5cm]{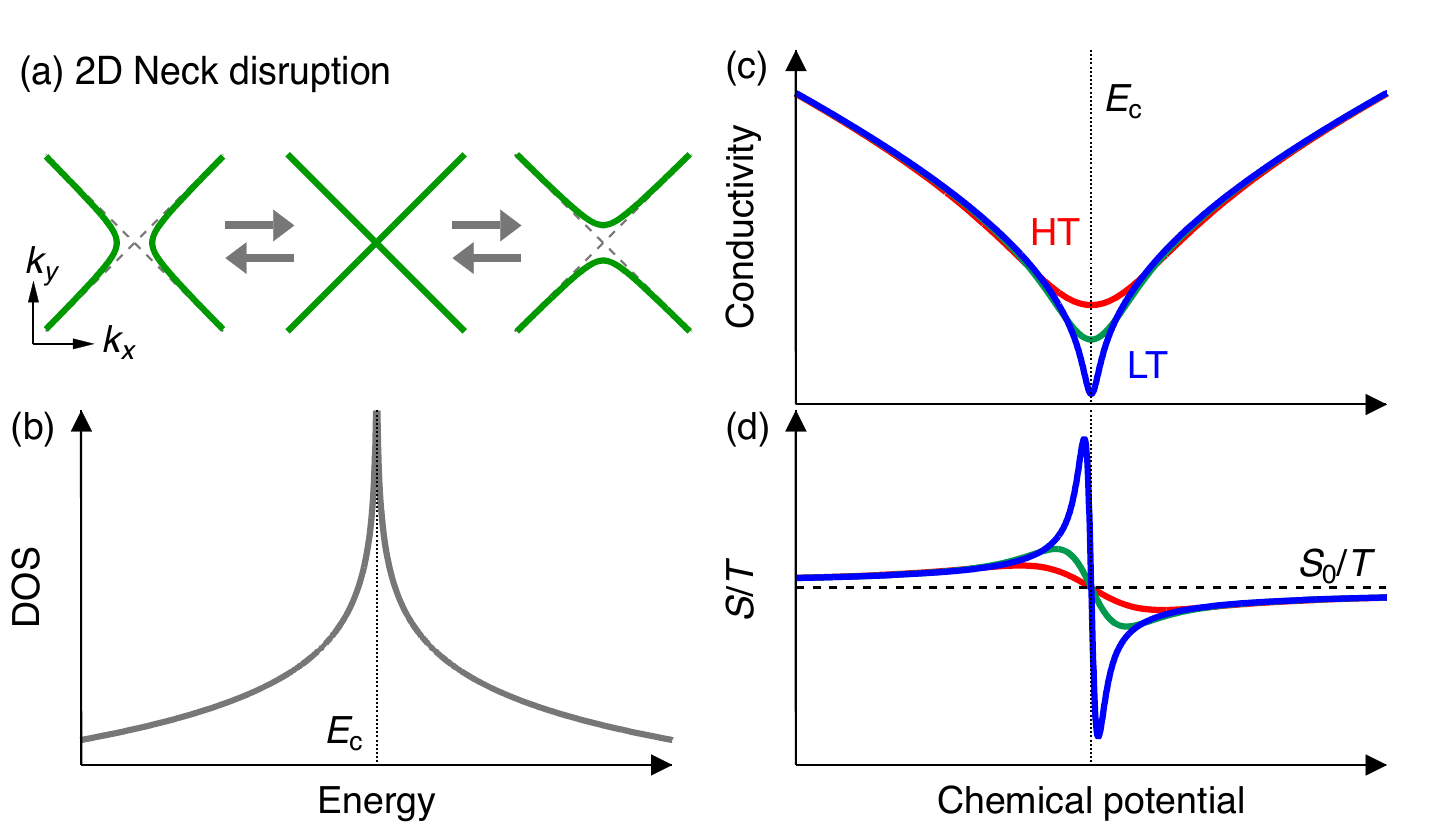}
\caption{
2D neck-disruption-type
Lifshitz transitions and the physical properties.
(a) The Fermi surface shape and (b) the DOS are symmetric around 
 the critical energy $E_{\rm c}$ at which 
the Lifshitz transition occurs.
(c) The conductivity and (d) the thermopower as a function of the chemical potential,
which corresponds to the amount of electron doping by La substitution,
for several temperatures. 
HT and LT represent high and low temperatures, respectively.
The vertical dotted lines show $E_{\rm c}$ and
the horizontal dashed line in (d) represents
the contributions from the regular part of the Fermi surfaces,
which is far from the VHS points.
}
\end{center}
\end{figure}

Here we discuss the energy dependence of the relaxation time $\tau(\varepsilon)$ near the Lifshitz transition.
Similar to the case of Sr$_2$RuO$_4$,
the 2D neck-disruption-type Lifshitz transition for the cylindrical Fermi surfaces [Fig. 3(a)]
has been investigated \cite{Blanter1992,Yamaji2006,Lin2010,Buhmann2013}.
In this case,
the energy dependence of the DOS $D(\varepsilon)$
shows logarithmic divergence near the critical energy $E_{\rm c}$ 
at which the Lifshitz transition occurs [Fig. 3(b)].
Such logarithmic DOS behavior is also confirmed by the numerical calculation for Sr$_2$RuO$_4$ [Fig. A2(a) in Appendix].
Then, through an elastic impurity scattering,
the scattering probability acquires a correction of 
the energy dependence of $1/\tau(\varepsilon)  \propto D(\varepsilon)$ \cite{Varlamov1989,Varlamov2021},
which significantly affects the energy
dependence of the conductivity function $\sigma(\varepsilon) \simeq D_0v_0^2\tau$.
Note that 
the DOS $D_0$ and the velocity $v_0$ in this conductivity function 
exhibit weak energy dependence because these mainly 
come from the electrons in the regular parts of the Fermi surfaces,
which are far from the VHS points \cite{Abrikosovbook}.
Figure 3(c) shows 
the calculated electrical conductivity for this simple model (Appendix).
The horizontal axis is the chemical potential and corresponds to the amount of electron doping by La substitution.
As a consequence, 
the electrical conductivity $\sigma$ decreases near $E_{\rm c}$
and 
such a modification
leads to a NFL-like resistivity of $\rho(T) = \rho_0 + AT^n$
with $n < 2$ for Sr$_2$RuO$_4$ \cite{Herman2019}, as 
observed near the VHS point for both La-substituted \cite{Kikugawa2004_La}
and uniaxially compressed \cite{Barber2018} cases.

In this model,
however,
as shown in Fig. 3(d), 
the thermopower should exhibit positive and negative peaks 
with the same magnitudes 
below and above $E_{\rm c}$, respectively \cite{Blanter1992,Yamaji2006,Lin2010},
as indicated from the Mott formula \cite{Behnia2004}
\begin{align}
\label{MOttf}
\frac{S}{T} 
\propto - \frac{1}{\sigma}\frac{\partial \sigma}{\partial \varepsilon}
\sim - \frac{1}{\tau}\frac{\partial \tau}{\partial \varepsilon},
\end{align}
where the energy dependence of $\tau$ is crucial as similar to 
the case of the conductivity as mentioned before.
It should be noted that 
such thermopower behavior with positive and negative peaks
is also obtained in the numerical calculations for Sr$_2$RuO$_4$ 
in the elastic impurity scattering regime \cite{Herman2019}.
In contrast,
for Sr$_{2-y}$La$_{y}$RuO$_4$,
the low-temperature thermopower seems to be enhanced only negatively near 
the critical content $y_{\rm c}\approx 0.2$ [Fig. 2(d)],
which is difficult to explain with 
such a conventional neck disruption case.
Also, as indicated in Fig. 2(d), 
an inelastic electron-electron scattering,
not included in the model of Fig. 3,
should be crucial 
near $y_{\rm c}$.

\begin{figure}[t]
\begin{center}
\includegraphics[width=8cm]{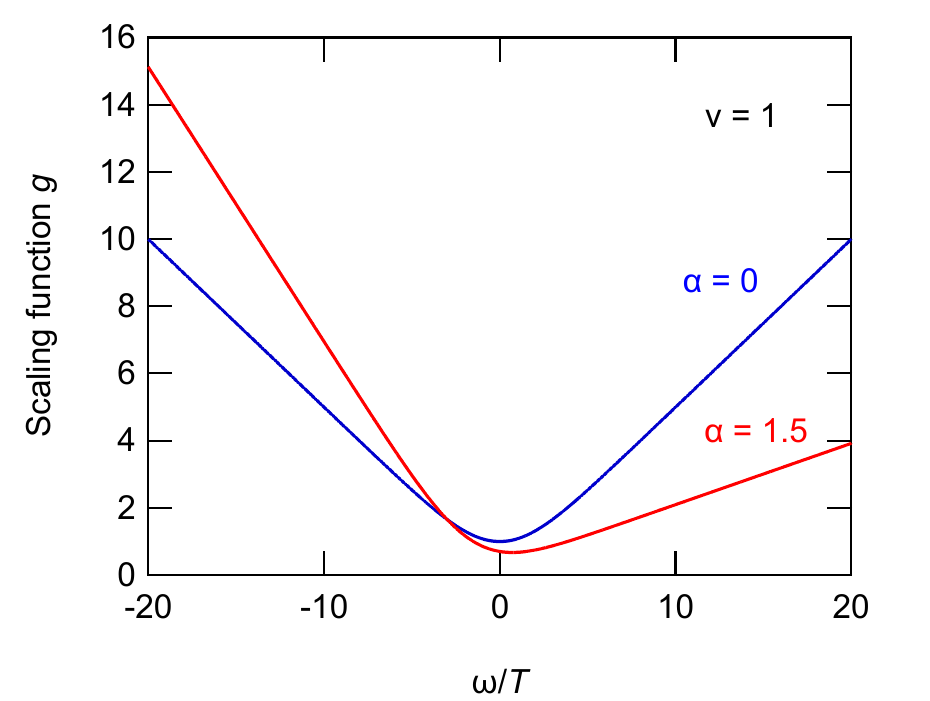}
\caption{
Scaling function $g(\omega/T)$
for an exponent $\nu = 1$ and an 
asymmetry parameter $\alpha = 0$ and $1.5$ \cite{Georges2021}.
The inelastic scattering rate 
$1/\tau_{\rm in}$
is 
given as $1/\tau_{\rm in} \propto (\pi T)^{\nu}g(\omega/T)$.
}
\end{center}
\end{figure}

To discuss the origin of the negatively enhanced thermopower near the VHS point,
we next consider a phenomenological model 
adopting a skewed NFL state \cite{Georges2021},
in which an asymmetric inelastic scattering rate $1/\tau_{\rm in} \propto (\pi T)^{\nu}g(\omega/T)$
characterized by a 
noneven scaling function $g(x) = |\Gamma(z)|^2\cosh(x/2)/[\cosh(\alpha/2)\Gamma\{(1+\nu)/2\}^2]$ becomes 
essential,
where 
$\Gamma(z)$ is the $\Gamma$ function,
$z = (1+\nu)/2 + i(x+\alpha)/2\pi$,
$\omega = \varepsilon - \mu$ is a relative energy from the chemical potential $\mu$,
\rc{$\nu$ $(\leq 1)$ is an exponent}, and 
$\alpha$ is an parameter to induce the asymmetry in the scattering rate.
Figure 4 shows the examples of a scaling function $g(\omega/T)$
with $\nu = 1$ for a symmetric (with an asymmetry parameter $\alpha = 0$) and
an asymmetric  ($\alpha = 1.5$) cases \cite{Georges2021}.
In the asymmetric case,
owing to the odd-frequency term of the scattering rate,
the contribution of either electrons or holes becomes stronger,
and most importantly,
the thermopower, a sensitive probe to the electron-hole asymmetry,
is enhanced either negatively or positively.
We infer that, along with the observation of the NFL resistivity near the VHS point \cite{Kikugawa2004_La},
the observed negative sign of the thermopower may be a hallmark of 
such a skewed NFL state.
Moreover, such unexpected sign change in the thermopower has also been 
discussed in the NFL regime of cuprate superconductors \cite{Gourgout2022}.
Note that the thermopower is also known as a sensitive probe for the entropic properties such as magnetic fluctuations but the present non-magnetic system may not be adapted to such situation.
At this stage, it is not easy to make a more quantitative calculation in this model, and
further theoretical study is necessary to quantitatively account for the the sign change.
Also, the frequency-dependent experiments such as the optical conductivity measurement will be crucial to examine the skewed NFL state.

Lifshitz transitions in correlated matter thus bring
an intriguing issue on the NFL state.
In this context,
Sr$_2$RuO$_4$,
in which a variety of electronic and magnetic states emerges under external perturbations  \cite{Nakatsuji2000,Minakata2001,Carlo2012,Ortmann2013,Zinkl2021},
is of peculiar interest  because 
it also exhibits the NFL behavior in Ti-substituted system \cite{Kikugawa2002Ti}.
Moreover,
recent experimental and theoretical studies have revealed the appearance of
enigmatic NFL states such as
a Planckian metal characterized by a linear temperature dependence of the resistivity \cite{Legros2019,Varma2020}
and the 
quantum critical phase in frustrated materials \cite{Zhao2019,Paschen2021},
deepening underlying physics of the NFL state of matter.

\section{summary}

In summary,
we have measured the thermopower of the electron-doped system Sr$_{2-y}$La$_{y}$RuO$_4$ 
and observed an unusual sign change in the thermopower near the Lifshitz transition at 
the VHS point $y_{\rm c}\approx 0.2$.
We discuss the thermopower in a skewed NFL state,
in which an asymmetric frequency dependence of the inelastic scattering rate is crucial,
\rc{possibly} leading to qualitative explanation for the negative sign of the thermopower near $y_{\rm c}$.
The present results thus 
offer a fascinating playground to investigate a variety of quantum phenomena of the correlated electrons near the Lifshitz transition.

\section*{Acknowledgments}

We appreciate Y. Fukumoto and R. Kurihara for discussion and R. Otsuki, H. Shiina, and R. Taira for the assistance.
This work was partly supported by JSPS KAKENHI Grant No. 17H06136, No. 21H01033, No. 22H01166, No. 22H01168, and No. 22K19093.

\section*{Appendix}

\renewcommand{\thefigure}{A\arabic{figure}}
\setcounter{figure}{0}

\subsection{First-principles calculations}

In order to investigate the thermopower theoretically, we
performed 
first-principles calculations based on density functional theory (DFT)
using Quantum Espresso \cite{Giannozzi2009,Giannozzi2017,Giannozzi2020}.
We used the projector-augmented-wave pseudopotentials 
with the Perdew-Burke-Ernzerhof generalized-gradient-approximation (PBE-GGA) exchange-correlation functional.
The cut-off energies for plane waves and charge densities were set to 
70 and 560 Ry, respectively,
and the $k$-point mesh was set to $20\times 20 \times 20$ uniform grid
to ensure the convergence.
Using the obtained eigenvalues of the $n$-th band at $\bm k$ point $E_{n,\bm k}$,
the DOS $D(\varepsilon) = \sum_{n,\bm k}\delta(\varepsilon-E_{n,\bm k})$ was obtained using the optimized tetrahedron method \cite{Kawamura2014},
where $\delta$ is the delta function.
We used on-site Coulomb energy $U = 3.5$~eV and
 exchange parameter $J = 0.6$~eV for Ru ions \cite{Huang2020}, and 
performed full relativistic calculations with spin-orbit coupling (${\rm DFT}+U+{\rm SOC}$).

\begin{figure}[t!]
\begin{center}
\includegraphics[width=8cm]{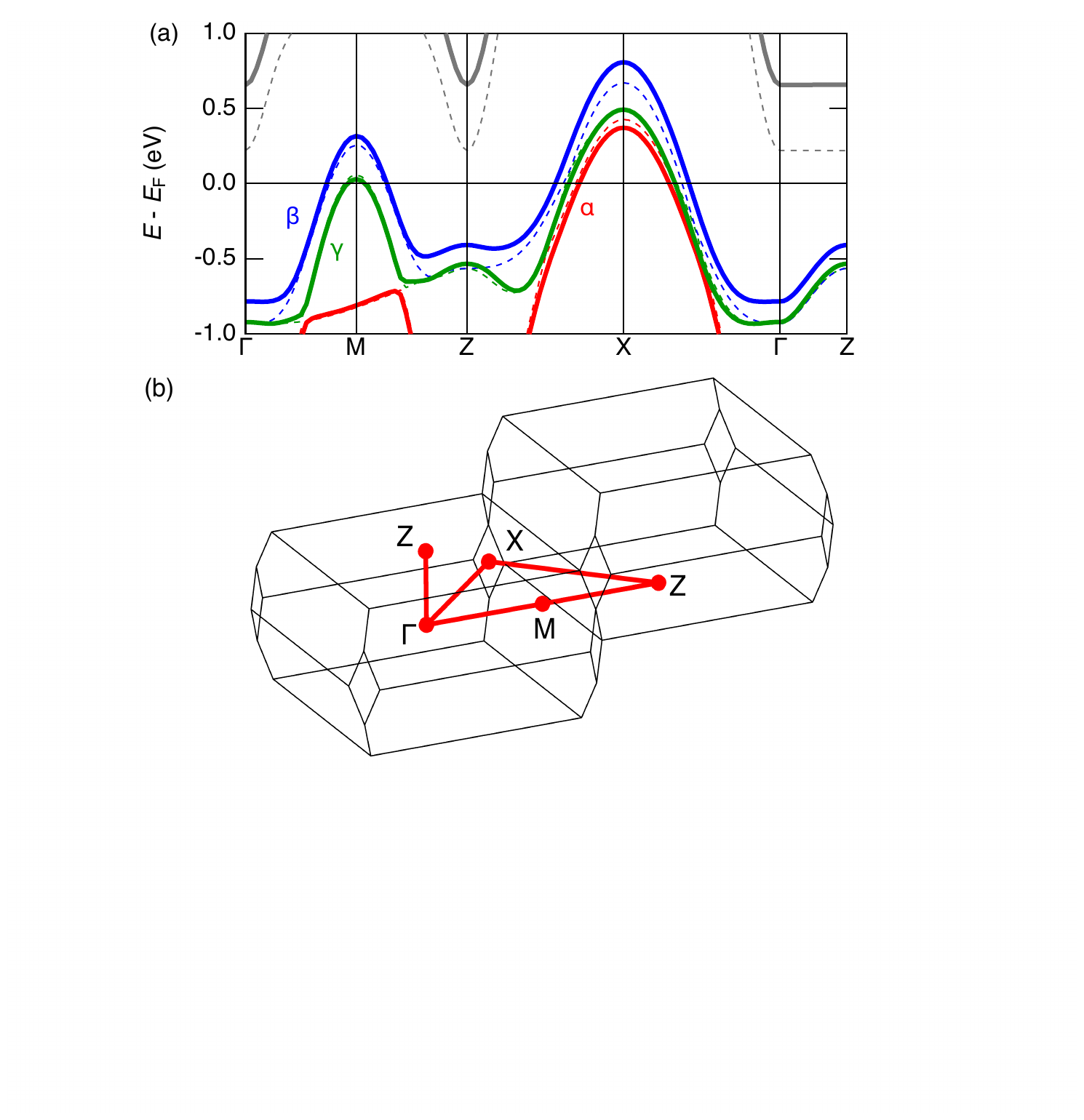}
\caption{
(a) Calculated band structure of Sr$_2$RuO$_4$ with ${\rm DFT}+U+{\rm SOC}$ scheme (solid curves).
The dashed curves represent the results of scalar relativistic calculations and $U$ is not included.
The Van Hove singularity is at the M point, at which the $\gamma$ band is slightly above $E_{\rm F}$.
(b) High-symmetry points in the Brillouin zone.
}
\end{center}
\end{figure}

Figure A1(a) shows the calculated electronic band structure near the Fermi energy $E_F$ of the parent material, 
which well coincides with the results in earlier studies \cite{Oguchi1995,Singh1995,Haverkort2008,Behrmann2012}.
The depicted $k$ path is shown in Fig. A1(b).
The $\beta$ and $\gamma$ bands at the high-symmetry points $\Gamma$ and Z split owing to 
the inclusion of SOC \cite{Haverkort2008}, and the $e_g$ bands are shifted upward due to the on-site $U$.
The calculated DOS is depicted in Fig. A2(a),
and
the Van Hove singularity (VHS) point of the  $\gamma$ band to show the cusp anomaly  is  shifted slightly from $E\approx50$~meV to
$E\approx30$~meV by including $U+{\rm SOC}$.
Such a trend is consistent with the results of ARPES experiment \cite{Shen2007,Tamai2019}.

\begin{figure}[t]
\begin{center}
\includegraphics[width=8cm]{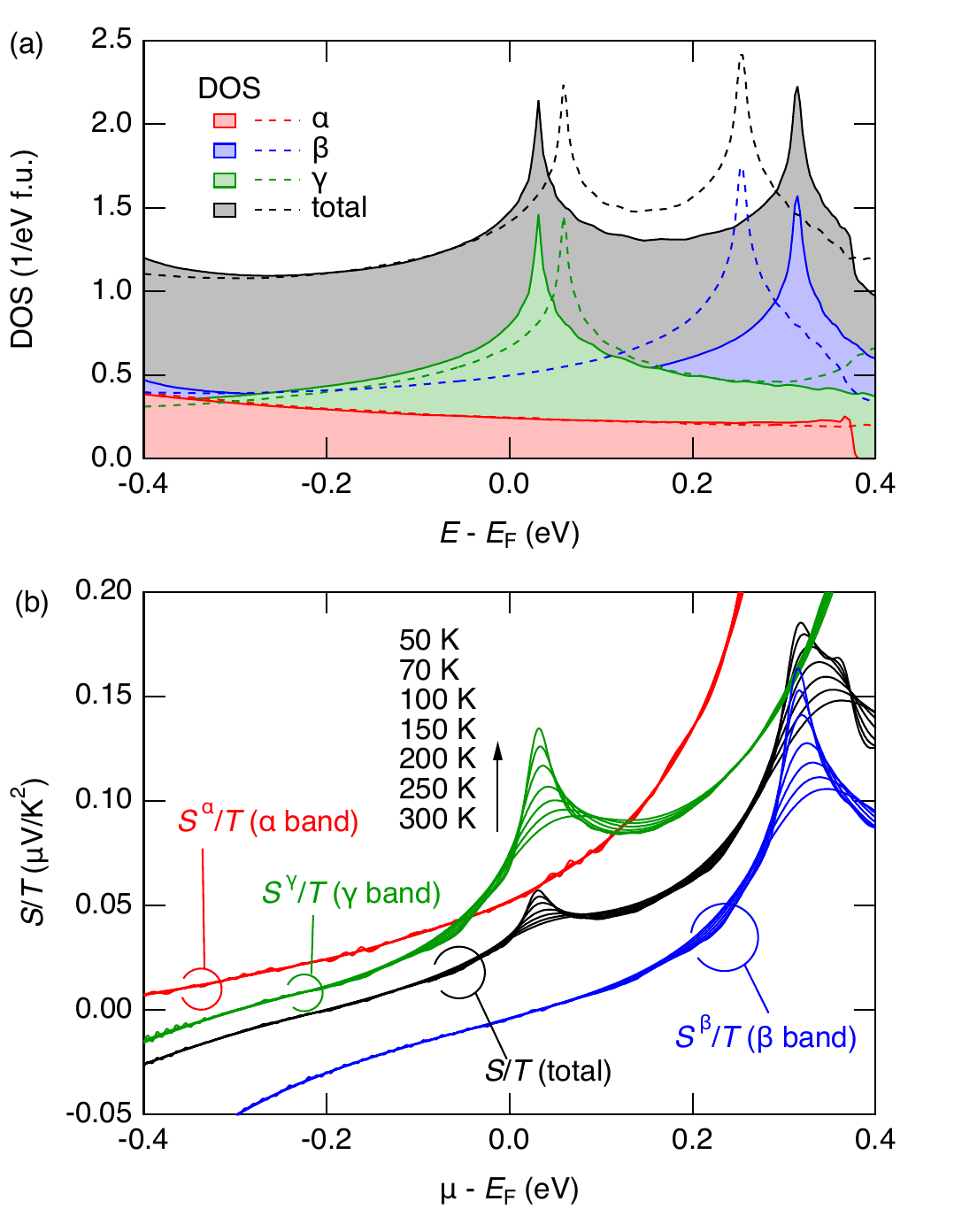}
\caption{
(a) Calculated total and partial density of states with ${\rm DFT}+U+{\rm SOC}$ scheme (solid filled curves).
The dashed curves represent the results of scalar relativistic calculations and $U$ is not included.
(b) Thermopower divided by temperature, $S/T$, calculated for several temperatures within the relaxation time approximation.
The horizontal axis shows the chemical potential measured from the Fermi energy of the parent compound.
The black curves represent the total $S/T$ and the red, blue, and green curves show the band-resolved
data $S^n/T$ for the $\alpha$, $\beta$, and $\gamma$ bands, respectively.
Varying $\mu$ across the VHS results in the enhancement of the positive $S/T$.
}
\end{center}
\end{figure}

\subsection{Calculated transport properties within a constant relaxation-time approximation}

To examine the experimentally observed singular behavior in $S/T$,
we have calculated the thermopower from the electronic band structure 
within the constant  relaxation time approximation.
Note that the correlation effect of $4d$ electrons is considerable in Sr$_2$RuO$_4$
and the calculation results with local-density approximation
are quantitatively different from the experimental observations \cite{Mravlje2016}.
On the other hand, 
the thermopower behavior in correlated metals is well described within a Fermi-liquid picture, 
where the correlation effect is included in the carrier effective mass \cite{Behnia2004}.
Also, in Sr$_{2-y}$La$_{y}$RuO$_4$,
the electron-doping effect by La substitution is well explained 
within a rigid band picture \cite{Kikugawa2004_La,Shen2007}.
We thus examine how the band structure affects the thermopower.

The transport coefficients were calculated based on the linearized Boltzmann equations 
under constant relaxation time approximation \cite{Madsen2006}.
The transport distribution function tensor $L_{ij}(\varepsilon)$ is calculated as
\begin{align}
L_{ij}(\varepsilon) = \sum_{n}L_{ij}^n(\varepsilon)
=
\sum_{n}\sum_{\bold k}v_iv_j\tau\delta(\varepsilon-E_{n,\bold k}),
\end{align}
where 
$L_{ij}^n(\varepsilon)$ is the partial transport distribution function tensor of the $n$-th band,
$v_i$ is the $i$-th component of the band velocity $\bold v = \frac{1}{\hbar}\nabla_{\bold k}E_{n,\bold k}$,
and
$\tau$ is the relaxation time.
We calculated the partial electrical conductivity tensor of the $n$-th band of 
$\sigma_{ij}^n(\mu) = e^2\int_{-\infty}^{\infty}d\varepsilon\left(-\frac{\partial f_0}{\partial \varepsilon}\right)L_{ij}^n$,
where $e$ is the elementary charge and $f_0$ is the Fermi-Dirac distribution function for 
the chemical potential $\mu$ and temperature $T$.
The total electrical conductivity tensor is 
$\sigma_{ij}(\mu) = \sum_{n}\sigma_{ij}^n(\mu)$.
Similarly, the partial Peltier conductivity tensor of $n$-th band $P_{ij}^n(\mu)=[\sigma S]_{ij}^n(\mu)$ is 
\begin{align}
P_{ij}^n(\mu) = -\frac{e}{T}\int_{-\infty}^{\infty}d\varepsilon\left(-\frac{\partial f_0}{\partial \varepsilon}\right)(\varepsilon-\mu)L_{ij}^n,
\end{align}
where $S_{ij}$ is the thermopower tensor.
The total Peltier conductivity is given as
$P_{ij}(\mu) = 
\sum_{n}P_{ij}^n(\mu)$.
Hereafter we consider the in-plane component ($ij=aa$) only and the subscript will be omitted.
The thermopower of the $n$-th band is then obtained as
$S^n=P^n/\sigma^n$,
and the total in-plane thermopower is given as
$S=P/\sigma=\sum_nP^n/\sum_n\sigma^n$
as is generally seen in multi-band systems.

Figure A2(b) depicts the thermopower divided by temperature, $S/T$, 
calculated for several temperatures within the constant relaxation time approximation.
The black curves represent the total $S/T$ and 
the red, blue, and green curves show the 
band-resolved $S^{n}/T$
for  $n=\alpha$, $\beta$, $\gamma$, respectively.
The horizontal axis is the chemical potential measured from the Fermi energy of the parent compound 
and corresponds to the amount of electron doping by La substitution.
Note that the calculated values of $S/T$ are significantly smaller than the experimental data because the 
electron correlation effect is not accurately included in this scheme and 
should be modified by using more precise methods such as the dynamical mean field theory \cite{Mravlje2016}.
Nevertheless, 
characteristic features reflecting the Lifshitz transition in Sr$_2$RuO$_4$ may be observed in the present calculations;
the band-resolved 
$S^{n}/T$ shows divergent behavior at low temperatures near $0.3$~eV ($\beta$ band) and 30 meV ($\gamma$ band)
corresponding to the VHS points of DOS for each band [Fig. A2(a)],
while  $S^{n}/T$ exhibits almost no temperature dependence for the $\alpha$ band like a simple metal.
In general, the VHS points of DOS strongly affect the thermopower \cite{Sato2023}.
It is now important that
the experimental data of $S/T$ seems to negatively diverge near the critical concentration [Fig. 2(d) in the main text],
while the calculated data is positively enhanced near the VHS points [Fig. A2(b)].
This discrepancy obviously originates from the energy dependence of the relaxation time 
ignored in the constant relaxation time approximation.

\subsection{Peculiar energy-dependent relaxation time and thermopower near the Lifshitz transition}

\begin{figure}[t]
\begin{center}
\includegraphics[width=8.5cm]{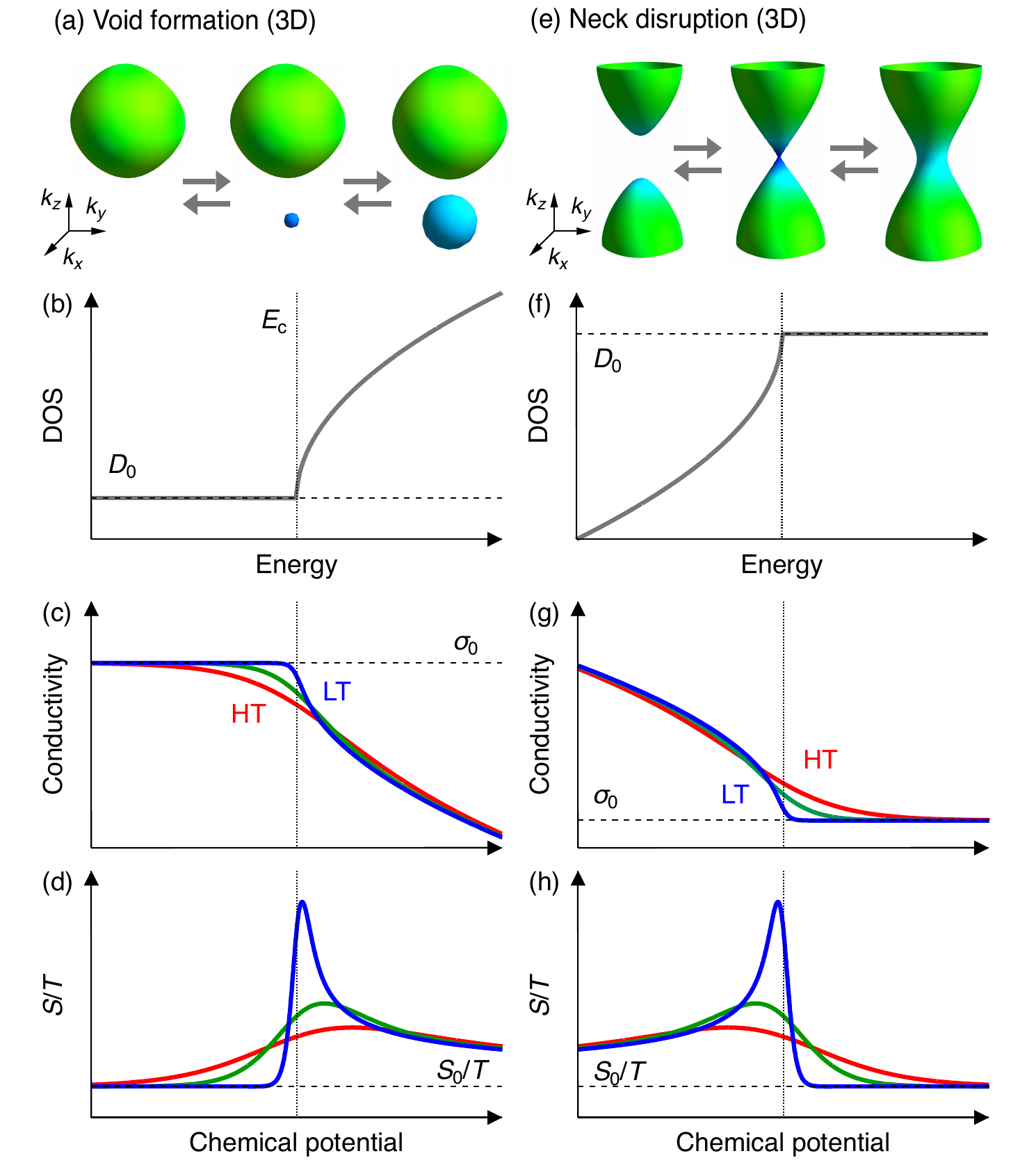}
\caption{
Two types of the Lifshitz transitions for 3D case and the physical properties:
(a-d) void formation and (e-h) neck disruption.
For each panel, 
the dashed lines represent 
the contributions from the regular part of the Fermi surfaces.
The vertical dotted lines show the critical energy $E_c$ at which 
the Lifshitz transition occurs.
HT and LT represent high and low temperatures, respectively.
}
\end{center}
\end{figure}

Here we briefly review the 
three-dimensional (3D) case to see the significance of the scattering process.
In the 3D case,
the Lifshitz transition is categorized into 
two types of topological change in the Fermi surface called void formation and neck disruption,
which are 
schematically shown in Figs. A3(a) and A3(e), respectively \cite{Lifshitz1960}.
In the void formation case, for example,
the DOS behaves as $D(\varepsilon) - D_0 \sim  |\varepsilon - E_c|^{1/2}$,
where $D_0$ is the DOS from the regular part of the Fermi surface (large void)
and
$E_c$ is the critical energy above which a new void appears [Fig. A3(b)].
At first glance,
such a small void with almost zero carrier velocity seems to give 
no contribution to the transport properties.
Through the scattering process,
however,
electrons in the regular part exchange the momenta 
with those in the singular part (small void)
and 
get virtually into
the singular part
\cite{Abrikosovbook,Varlamov1989,Varlamov2021}.
As a result, according to the golden rule, 
the scattering probability acquires a correction of 
the energy dependence of $1/\tau(\varepsilon)  \propto D(\varepsilon)$,
and then 
the electrical conductivity $\sigma \propto \tau$
decreases above $E_c$ [Fig. A3(c)]. 
Note that the energy dependence of $\tau$ is essential here
as
an approximate form
$\sigma(\mu)
\simeq
e^2\int_{-\infty}^{\infty}d\varepsilon\left(-\frac{\partial f_0}{\partial \varepsilon}\right)Dv^2\tau
\simeq e^2 D_0v_0^2\tau(\mu)$,
because 
the DOS $D_0$ and the velocity $v_0$ in the conductivity
mainly come from the electrons in the large void \cite{Abrikosovbook}.

According to the Mott formula,
the thermopower $S$ is given as 
\begin{align}
\frac{S}{T} 
\propto - \frac{1}{\sigma}\frac{\partial \sigma}{\partial \varepsilon}
\sim - \frac{1}{\tau}\frac{\partial \tau}{\partial \varepsilon},
\end{align}
where the energy dependence of $\tau$ is crucial as similar to 
the case of the conductivity,
and thus it shows a sharp peak structure 
at low temperatures [Fig. A3(d)].
The similar situation occurs in the case of the 
neck disruption [Figs. A3(e-h)] and the 
thermopower is also enhanced positively near the critical point.
Note that 
the singularities in $\sigma$ and $S$ are smeared with increasing temperature.
These thermoelectric singularities in 3D case have been experimentally observed in the 
Li-Mg alloy \cite{Egorov1983}.

\subsection{Calculations of the transport properties for the Lifshitz transitions}

Here we  show the calculation details for the transport coefficients near the Lifshitz transition by using the 
energy-dependent scattering time.
The electrical conductivity $\sigma$ and the Peltier conductivity $P=\sigma S$ are given as
\begin{align}
\begin{bmatrix}
\sigma \\
P \\
\end{bmatrix}
&=
\begin{bmatrix}
e^2 \\
-\frac{e}{T} \\
\end{bmatrix}
\int_{-\infty}^{\infty}
d\varepsilon
\left(-\frac{\partial f_0}{\partial \varepsilon}\right)
\begin{bmatrix}
1 \\
\varepsilon -\mu \\
\end{bmatrix}
L
\\
&=
\frac{e}{4k_BT^2}
\int_{-\infty}^{\infty}
\frac{d\omega}{\cosh^2 (\beta\omega/2)}
\begin{bmatrix}
eT \\
-\omega \\
\end{bmatrix}
L,
\end{align}
where 
$\omega = \varepsilon -\mu$ is the relative energy measured from the chemical potential.
Using the energy-dependent scattering time,
the transport function 
near the Lifshitz transition is approximately given as
\begin{align}
L = 
D_0v_0^2\tau(\varepsilon),
\end{align}
and the transport coefficients are given as
\begin{align}
\begin{bmatrix}
\sigma \\
P \\
\end{bmatrix}
&=
\frac{eD_0v_0^2}{4k_BT^2}
\int_{-\infty}^{\infty}
\frac{d\omega}{\cosh^2 (\beta\omega/2)}
\begin{bmatrix}
eT \\
-\omega \\
\end{bmatrix}
\tau(\omega),
\end{align}
where the scattering time is model-dependent as described below.

For the 3D void formation case in Figs. A3(a-d), 
the density of states is expressed as
\begin{align}
D(\varepsilon)
&\sim 
D_0
+a|\varepsilon-E_c|^{1/2}\theta(\varepsilon-E_c),
\end{align}
where $a$ $(>0)$ is a constant
and $\theta$ is the Heaviside step function,
as is shown in Fig. A3(b).
The scattering time is then given as
\begin{align}
\tau(\varepsilon)
&\sim D(\varepsilon)^{-1}\\
&\sim 
D_0
-a|\varepsilon-E_c|^{1/2}\theta(\varepsilon-E_c)\\
&=
D_0
-a|\omega+Z|^{1/2}\theta(\omega+Z),
\end{align}
where $Z = \mu - E_c$ is the chemical potential measured from the critical energy.
The transport coefficients are now calculated and the 
thermopower $S$ is give as $S = P/\sigma$.
Note that although the regular part $D_0$ also depends on the energy \cite{Varlamov1989,Varlamov2021},
the energy dependence of the singular part is much significant.
Indeed, the calculation results shown in Fig. A3 are similar to the earlier works.

For the 3D neck disruption case [Figs. A3(e-h)],
the scattering time is given as
\begin{align}
\tau(\varepsilon)
&\sim 
D_0
+a|E_c-\varepsilon|^{1/2}\theta(E_c-\varepsilon)\\
&=
D_0
+a|-\omega-Z|^{1/2}\theta(-\omega-Z),
\end{align}
which is similar to the case of void formation.

For the symmetric 2D neck disruption case  [Figs. 3(a-d) in the main text], 
the density of states near the Lifshitz transition is given as
a logarithmic form of 
\begin{align}
D(\varepsilon)
&\sim 
\ln\frac{t}{|\varepsilon-E_c|},
\end{align}
where $t$ $(>0)$ is a constant as is shown in Fig. 3(b) in the main text.
The scattering time is given as
\begin{align}
\tau(\varepsilon)
&\sim 
\left(\ln\frac{t}{|\varepsilon-E_c|}\right)^{-1}
=\left(\ln\frac{t}{|\omega + Z|}\right)^{-1}.
\end{align}


\end{document}